\documentclass[11pt,a4paper]{article}
\usepackage{moriond}
\usepackage{graphicx}
\RequirePackage{xspace}

\def\Journal#1#2#3#4{{#1} {\bf #2}, #3 (#4)}

\def\NIMA{{\em Nucl. Instrum. Methods} A}

\def\PLB{{\em Phys. Lett.}  B}
\def\PRL{\em Phys. Rev. Lett.}
\def\PRD{{\em Phys. Rev.} D}
\def\PRDRC{{\em Phys. Rev.} D {\em (Rapid Communications)}}

\def\ra{\rightarrow}

\def\be{\begin{equation}}
\def\ee{\end{equation}}
\def\bea{\begin{eqnarray}}
\def\eea{\end{eqnarray}}

\usepackage{relsize}
\def\babar{\mbox{\slshape B\kern-0.1em{\smaller A}\kern-0.1em
    B\kern-0.1em{\smaller A\kern-0.2em R}}}
\newcommand{\mevcc}{\ensuremath{{\mathrm{\,Me\kern -0.1em V\!/}c^2}}\xspace}
\newcommand{\mev}{\ensuremath{\mathrm{\,Me\kern -0.1em V}}\xspace}

\begin{document}
\vspace*{4cm}
\title{QUARKONIUM AND CHARM HADRONS : NEW RESULTS ON SPECTROSCOPY IN BABAR}

\author{ S.J. GOWDY FOR THE \babar\ COLLABORATION }

\address{Stanford Linear Accelerator Center, 2575 Sand Hill Rd, Menlo
Park CA 94025, USA}

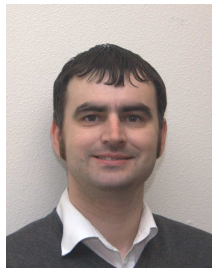
\maketitle

\abstracts{We report on recent results in spectroscopy from
\babar. This includes the discovery of a new $\Lambda_c$ baryon,
detailed studies of the $\mathrm{D}^*_{sJ}(2317)^+$,
$\mathrm{D}_{sJ}(2460)^+$, X(3872) and Y(4260) particles and the first
measurement of hadronic non-$\mathrm{B\overline{B}}$ decays of the
$\Upsilon(4\mathrm{S})$ meson.}

\section{\babar}

The \babar\ detector is described in detail in a prior publication
\cite{babar:02}.

\section{Charm Baryon Spectroscopy}

The current particle assignment by the Particle Data Group
\cite{eid:04} is assigned indirectly rather than being based on
angular analysis. The $\Lambda_c(2785)$ and $\Lambda_c(2880)$ states
\cite{cleo:01} are only listed as single or double asterisk
(respectively) as they have not yet been confirmed.


In studying the $\mathrm{D}^0\mathrm{p}$ spectra
\babar\ has observed two charm baryons\cite{bbr:06_1}, the
$\Lambda_c(2880)$ and the $\Lambda_c(2940)$. The data sample in this
study was 288 fb$^{-1}$. The $\mathrm{D}^0$ was found from its decay
to either $\mathrm{K}^-\pi^+$ or $\mathrm{K}^-\pi^+\pi^-\pi^+$. No
signal for these particles was seen in the $\mathrm{D}^+\mathrm{p}$
mode which rules out a $\Sigma_c$ states.




The observation of these particles in this channel rules out some
other non-$\Lambda_c$ interpretations\cite{blech:03};
$\Sigma_{c2}(\frac{3}{2},\frac{5}{2})$,
$\Lambda^\prime_{c1}(\frac{3}{2})$ and
$\Lambda^\prime_{c2}(\frac{3}{2},\frac{5}{2})$.

The first of these is the $\Lambda_c(2880)$, confirming CLEO's
discovery \cite{cleo:01}. The particle's parameters found in this
analysis are;

\begin{displaymath}
\renewcommand{\arraystretch}{1.2}
\begin{array}{r@{\;}c@{\;}r@{\;}c@{\;}r@{\;}l@{\;}c@{\;}r@{\;}l@{\;}l}
m &=[& 
      2881.9 &\pm& 0.1 &\mathrm{(stat.)}& \pm& 0.5 &\mathrm{(syst.)}& ]~\mevcc \\
\Gamma &=[& 
         5.8 &\pm& 1.5 &\mathrm{(stat.)}& \pm& 1.1 &\mathrm{(syst.)}& ]~\mev\;.
\end{array}
\end{displaymath}

\newpage

A new baryon has been discovered with the properties;

\begin{displaymath}
\renewcommand{\arraystretch}{1.2}
\begin{array}{r@{\;}c@{\;}r@{\;}c@{\;}r@{\;}l@{\;}c@{\;}r@{\;}l@{\;}l}
m &=[& 
      2939.8 &\pm& 1.3 &\mathrm{(stat.)}& \pm& 1.0 &\mathrm{(syst.)}& ]~\mevcc \\
\Gamma &=[& 
        17.5 &\pm& 5.2 &\mathrm{(stat.)}& \pm& 5.9 &\mathrm{(syst.)}& ]~\mev\;.
\end{array}
\end{displaymath}

This state has been given the name $\Lambda_c(2940)^+$. There is no
evidence of doubly-charged partners in the $\mathrm{D}^+\mathrm{p}$
spectrum.

\section{Charm Meson Spectroscopy}

Current theory of treating $\mathrm{c\overline{s}}$ system like the
hydrogen atom (with one heavy and one light element) has been fairly
successful. Prior to 2003 this theory had postulated many particles
that were later observed as expected. A few expected states were not
observed and this was ascribed to their predicted large widths.


The $\mathrm{D}^*_{sJ}(2317)^+$ was discovered by
\babar\cite{bbr:03}. CLEO published\cite{cleo:03} the discovery of the
$\mathrm{D}_{sJ}(2460)^+$.

An extensive study of decays to $\mathrm{D}_s^+$ plus one or two
particles of the types $\pi\pm$, $\pi^0$ or $\gamma$ has been carried
out\cite{bbr:06_2} looking at 232fb$^{-1}$ of data.

The $\mathrm{D}_s^+$ was reconstructed from only
$\mathrm{K}^+\mathrm{K}^-\pi^+$ decays. Furthermore to reduce
background, only certain intermediate states were allowed: either a
$\phi(1020)$ could be reconstructed from the
$\mathrm{K}^+\mathrm{K}^-$ or a $\overline{\mathrm{K}}^*(892)$ from
$\mathrm{K}^-\pi^+$.




Only three decay modes are measured (the rest only have upper limits
reported in this analysis). The ratio of the measured branching
fractions are;

\begin{displaymath}
\renewcommand{\arraystretch}{1.8}
\begin{array}{ccl@{}r@{\:}c@{\:}r@{\:}c@{\:}r}
\frac{\mathcal B(\mathrm{D}_{sJ}(2460)^+\to\mathrm{D}_s^+\gamma)}{\mathcal B(\mathrm{D}_{sJ}(2460)^+\to\mathrm{D}_s^+\pi^0\gamma)}
&=& &0.337 &\pm&   0.036  &\pm&   0.038 \\
\frac{\mathcal B(\mathrm{D}_{sJ}(2460)^+\to\mathrm{D}_s^+\pi^+\pi^-)}{\mathcal B(\mathrm{D}_{sJ}(2460)^+\to\mathrm{D}_s^+\pi^0\gamma)}
&=& &0.077 &\pm&   0.013  &\pm&    0.008
\end{array}
\end{displaymath}

The decays seen are consistent with both particles being P-wave
c$\overline{\mathrm{s}}$ mesons with the $\mathrm{D}^*_{sJ}(2317)^+$
being $0^+$ and the $\mathrm{D}_{sJ}(2460)^+$ being $1^+$ states. The
particle properties measured in this analysis are shown in Table
\ref{tab:Ds}. It should be noted that no intrinsic width is measured.

\begin{table}

\caption{\label{tab:Ds}A summary of the combined mass and width
results. For the masses, the first quoted uncertainty is statistical
and the second is systematic. The limits on the intrinsic width
$\Gamma$ are at 95\% CL.}

\begin{center}
\begin{tabular}{l@{\hspace{0.5cm}}r@{$\:\pm\:$}r@{$\:\pm\:$}r@{\hspace{0.5cm}}r}
\hline\hline
Particle & \multicolumn{3}{c}{Mass (\mevcc)} & $\Gamma$ (\mev) \\
\hline
$\mathrm{D}^*_{sJ}(2317)^+$ & 2319.6 & 0.2 & 1.4 & $<3.8$ \\
$\mathrm{D}_{sJ}(2460)^+$ & 2460.1 & 0.2 & 0.8 & $<3.5$ \\
$\mathrm{D}_{s1}(2536)^+$ & 2534.6 & 0.3 & 0.7 & $<2.5$ \\
\hline\hline
\end{tabular}

\end{center}

\end{table}

No evidence was found for doubly-charged or neutral partners of the
$\mathrm{D}^*_{sJ}(2317)^+$. This deficit disfavours some molecule
interpretations of these states.

The fully reconstructed B sample was used to also study twelve decay
modes of neutral and charged B mesons of the form
$\mathrm{D}_{s(J)}^{(*)}\mathrm{D}^{(*)}$. In this sample the
properties of the the B mesons are well known due to all decay
products of one of the B mesons being measured. This analysis used 230
million $\mathrm{B\overline{B}}$ pairs\cite{bbr:06_3}.

All decay products of one of the D mesons are also measured, the other
D could decay to anything and is represented as $\mathrm{D}_X$, its
mass and momenta inferred by kinematics. The mass spectra of the
$\mathrm{D}_X$ system is studied in each of the modes to determine
absolute branching fractions.




Using previously published\cite{bbr:04} \babar\ study of the branching
fraction for $\mathrm{B}\to\mathrm{D}_{sJ}(2460)^+\mathrm{D}^{(*)}$ it
is possible to determine some absolute branching fractions;

\begin{displaymath}
\begin{array}{rlcrcrcrr}
B(\mathrm{D}_{sJ}(2460)^+ &\ra \mathrm{D}^{*+}_s \pi^0) &=& (56 &\pm& 13_{\rm stat.} &\pm& 9_{\rm syst.})&\%  \\ 
B(\mathrm{D}_{sJ}(2460)^+ &\ra \mathrm{D}_s^+ \gamma)  &=& (16 &\pm& 4_{\rm stat.} &\pm& 3_{\rm syst.})&\% \\
\end{array}
\end{displaymath}

We therefore know that the $\mathrm{D}_{sJ}(2460)^+$ decays via a
photon or a $\pi^0$ emission to a $\mathrm{D}_s^{(*)+}$ $(72\pm19)$\%
of the time. Also known from branching fraction ratios reported
earlier in this section together with this result that approximately
another 4\% emit a pair of charged pions while decaying to a
$\mathrm{D}_s^+$.

\section{Charmonium Spectroscopy}

There are many models for Charmonium spectroscopy that predict as yet
unseen particles\cite{bar:05}. Some of these may be the recently
observed X(3872) and Y(4260) states.


The X(3872) particle was discovered by Belle\cite{bel:03} and
confirmed by \babar\cite{bbr:05}. It is unknown if it is a
$\mathrm{c\overline{c}}$ pair, diquark-antidiquark pair or a
$\mathrm{D}^0\mathrm{\overline{D}}^{*0}$ molecule.

The diquark model would predict that the X(3872) is actually two
amplitudes of a nonet, one visible via charged and other via neutral B
mesons decays. The predicted mass difference of the $\mathrm{X}_u$ and
$\mathrm{X}_d$ states is $7\pm2\mevcc$. The difference has been
measured\cite{bbr:06_4} to be $2.7\pm1.3\pm0.2\mevcc$. This is
consistent with both the prediction and zero.

Some molecules models predict that the ration of these branching
fractions should be less than 10\%. The measurement shows the ratio to
be between 0.13 and 1.10 at the 90\% confidence level, so those models
are disfavoured.

Belle has reported evidence\cite{bel:05} of the decay
X(3872)$\to\mathrm{J}/\psi\gamma$ in B meson decays. An analysis has
been done using 287 million $\mathrm{B\overline{B}}$ decays to confirm
this. The following branching fraction was also measured to validate
signal extraction method and is found to agree with the the Particle
Data Group value;

\begin{displaymath}
B(\mathrm{B}^+\to\chi_{c1}\mathrm{K}^+) = (5.6\pm0.2\pm0.7) \times 10^{-4}
\end{displaymath}

The analysis finds $19.2\pm5.7$ events and a product branching
fraction of;

\begin{displaymath}
B(\mathrm{B}^+\to\mathrm{X}(3872)\mathrm{K}^+,\mathrm{X}(3872)\to\mathrm{J}/\psi\gamma) = (3.4\pm1.0\pm0.3) \times 10^{-6}
\end{displaymath}

The observation of this mode implies that the X(3872) is a C=+1
state. The partial width ratio has also been measured to be;

\begin{displaymath}
\frac{\Gamma(\mathrm{X}(3872)\to\mathrm{J}/\psi\gamma)}
{\Gamma(\mathrm{X}(3872)\to\mathrm{J}/\psi\pi^+\pi^-)} = 0.34\pm0.14
\end{displaymath}

232fb$^{-1}$ of initial state radiation (ISR) data has been studied
looking for the X(3872) but no evidence was found.
This unsuccessful search for X(3872) 
resulted in the discovery of the Y(4260)\cite{bbr:05_2}. This state
has been confirmed by CLEO\cite{cleo:06}.

A study has been carried out on 232 million $\mathrm{B\overline{B}}$
decays\cite{bbr:06_4}. This study shows a consistent feature with the
discovery spectra and measures a product branching fraction of;

\begin{displaymath}
B(\mathrm{B}^-\to\mathrm{Y}(4260)\mathrm{K}^-,\mathrm{Y}(4260)
				\to\mathrm{J}/\psi\pi^+\pi^-)
= (2.0\pm0.7\pm0.2) \times 10^{-5}
\end{displaymath}

An analysis of 232fb$^{-1}$ of ISR data using the final state
$\mathrm{K^+K^-}\pi^+\pi^-$ has been carried out, in this analysis the
ISR photon was observed. Events are selected where the kaon pair comes
from a $\phi$ to provide a confidence limit (CL);

\begin{displaymath}
B(\mathrm{Y}(4260)\to\phi\pi^+\pi^-).\Gamma^\mathrm{Y}_{ee} < 0.4 \mathrm{eV}\hspace{2mm}\mathrm{at} \hspace{2mm} 90\% \hspace{2mm}\mathrm{CL}
\end{displaymath}

A recent \babar\ result\cite{bbr:05_2} can be use to obtain the ratio
of branching fractions;

\begin{displaymath}
\frac{B(\mathrm{Y}(4260)\to\phi\pi^+\pi^-)}{B(\mathrm{Y}(4260)\to\mathrm{J}/\psi\pi^+\pi^-)}
< 0.102 \hspace{2mm}\mathrm{at} \hspace{2mm} 90\% \hspace{2mm}\mathrm{CL}
\end{displaymath}

The glueball hypothesis predicts\cite{zhu:05} these to be equal,
therefore it is disfavoured.

\section{Bottomonium Spectroscopy}

Decays of the $\Upsilon$ states above the $\mathrm{B\overline{B}}$
have been primarily to that state. \babar\ has now measured some
non-$\mathrm{B\overline{B}}$ hadronic decays of the
$\Upsilon(4\mathrm{S})$.


This analysis\cite{bbr:06_5} used $(230.0\pm2.5)\times 10^6
\Upsilon(4\mathrm{S})$ decays, which corresponds to 211fb$^{-1}$ of
on-peak data. As a cross check another 22fb$^{-1}$ of off-peak data
(approximately 40MeV below the $\Upsilon(4\mathrm{S})$ peak) was
searched for for any indications of $\Upsilon(4\mathrm{S})$ decays and
none were found.

The decays being searched for are
$\Upsilon(4\mathrm{S})\to\Upsilon(n\mathrm{S})\pi^+\pi^-$, with
n=1,2. The $\Upsilon(n\mathrm{S})$ states are reconstructed from
oppositely charged pairs of electrons or muons. However, due to high
background from Bhabha events only the muon decays are used in the
measurement. The electron sample is use as a cross check.

The branching fractions obtained by using the Particle Data Group's
averages for $B(\Upsilon(n\mathrm{S})\to\mu^+\mu^-)$;

\begin{displaymath}
\begin{array}{rcrcrr}
B(\Upsilon(4\mathrm{S})\to\Upsilon(1\mathrm{S})\pi^+\pi^-)&=&(0.90&\pm&0.15)&\times10^{-4}\\
B(\Upsilon(4\mathrm{S})\to\Upsilon(2\mathrm{S})\pi^+\pi^-)&=&(1.29&\pm&0.32)&\times10^{-4}\\
\end{array}
\end{displaymath}

\section{Summary}

A new charmed baryon has been observed. In addition, for the first
time a decay of a charm baryon to a non-charm baryon has been
measured. Many enhancements to the knowledge of the
$\mathrm{D}^*_{sJ}(2317)^+$, $\mathrm{D}_{sJ}(2460)^+$, X(3872) and
Y(4260) particles have been accomplished. We have reported on the first
measurement of hadronic non-$\mathrm{B\overline{B}}$ decays of the
$\Upsilon(4\mathrm{S})$ meson.

\end{document}